\documentclass[12pt]{iopart}

\usepackage{iopams}  

\usepackage{lineno}

\usepackage{graphicx}
\usepackage{subfigure}

\usepackage[authoryear]{natbib} 
\usepackage{threeparttable}
\usepackage{color}

\begin{document}

\title[Mathematical analysis of FLASH effect models]{Mathematical Analysis of FLASH Effect Models Based on Theoretical Hypotheses}

\author{Ankang Hu$^{1,2}$, Wanyi Zhou$^{1,2}$, Rui Qiu$^{1,2,*}$ and Junli Li$^{1,2,*}$}

\address{$^1$Department of Engineering Physics, Tsinghua University, Beijing, China\\
$^2$Key Laboratory of Particle and Radiation Imaging, Tsinghua University, Ministry of Education, Beijing, China\\
$^*$Authors to whom any correspondence should be addressed}
\ead{lijunli@mail.tsinghua.edu.cn and qiurui@tsinghua.edu.cn}
\vspace{10pt}
\begin{indented}
\item[]November 2023
\end{indented}

\begin{abstract}

Objective: Clinical applications of FLASH radiotherapy require formulas to describe how the FLASH radiation features and other related factors determine the FLASH effect. Mathematical analysis of the models can connect the theoretical hypotheses with the radiobiological effect, which provides the foundation for establishing clinical application models. Moreover, experimental and clinical data can be used to explore the key factors through mathematical analysis. 

Approach: We abstract the complex models of the oxygen depletion hypothesis and radical recombination-antioxidants hypothesis into concise mathematical equations. The equations are solved to analyze how the radiation features and other factors influence the FLASH effect. Then we propose methodologies for determining the parameters in the models and utilizing the models to predict the FLASH effect.

Main results: The formulas linking the physical, chemical and biological factors to the FLASH effect are obtained through mathematical derivation of the equation. The analysis indicates that the initial oxygen concentration, radiolytic oxygen consumption and oxygen recovery are key factors for the oxygen depletion hypothesis and that the level of antioxidants is the key factor for the radical recombination-antioxidants hypothesis. According to the model derivations and analysis, the methodologies for determining parameters and predicting the FLASH effect are proposed: (1)the criteria for data filtration, (2)the strategy of hybrid FLASH and conventional dose rate (CONV) irradiation to ensure the acquisition of effective experimental data across a wide dose range, (3)the pipelines of fitting parameters and predicting the FLASH effect.

Significance: This study establishes the quantitative relationship between the FLASH effect and key factors. The derived formulas can be used to calculate the FLASH effect in future clinical FLASH radiotherapy. The proposed methodologies guide to obtain sufficient high-quality datasets and utilize them to predict the FLASH effect. Furthermore, this study indicates the key factors of the FLASH effect and offers clues to further explore the FLASH mechanism.

\end{abstract}


\maketitle

\section{Introduction}
\indent FLASH radiotherapy has emerged as a pioneering area of interest in the radiotherapy domain, garthering significant attention due to its unique radiobiological properties. This approach has demonstrated the potential to preserve normal tissue while maintaining tumor control comparable to that achieved with conventional dose rate (CONV) radiotherapy, thereby enhancing the efficacy of cancer treatment and offering the prospect of improved outcomes for a multitude of cancer patients. 
The FLASH effect has been documented in preclinical studies using a variety of radiation modalities, including electrons \citep{favaudon_2014}, low-energy X-rays \citep{MontatGruel_2018}, high-energy X-rays \citep{Gao_2021}, protons \citep{Rama_2019}, and even carbon ions \citep{Tinganelli_2022}. The clinical application of FLASH radiotherapy is currently being explored in ongoing trials \citep{Mascia_2022}.

However, the mechanism of FLASH effect remains unclear, inspiring diligent efforts to unravel its complexities. Several analytical hypotheses have been introduced. 
The oxygen depletion hypothesis \citep{Pratx_2019,ZhuH_2021, Zou_2022} is one of the most widely discussed hypotheses. 
It postulates that the high dose delivered within a short time instantly consumes a large amount of oxygen and induces the tissue hypoxic because of the limited oxygen recovery speed. Thus, the hypoxic tissue can exhibit radioresistance, resulting in the tissue sparing effect.
However, the hypothesis faces the challenge of explaining the equivalent tumor control.
The radical recombination hypothesis \citep{Labarbe_2022} and its expansion, radical recombination-antioxidants hypothesis \citep{hu_radical_2023}, attempt to explain the FLASH effect mechanism from the perspective of radiochemistry.
It postulates that a high dose in a short time generates a high transient concentration of peroxyl radicals, leading to a higher portion of radical recombination. The radical recombination reaction generates non-radical products and reduces radical-induced damage. The antioxidants compete to react with peroxyl radicals. The concentration of antioxidants in tumors is generally several times higher than that of normal tissues. 
Thus, the portion of radical recombination in high-antioxidant tissue remains small under both FLASH and CONV irradiation, with the resulting damage being indistinguishable, which can explain the comparable tumor control observed in the FLASH effect experiments.
Besides, some researchers proposed their hypothesis based on their theoretical models or experimental results, such as “protection of circulating immune cells” \citep{Jin_2019}, “DNA integrity” \citep{Shi_2022} and “mitochondrial damage response” \citep{Guo_2022}. These hypotheses provide diverse explanations for the FLASH effect, contributing valuable references for future investigations. While some hypotheses qualitatively propose potential mechanisms, others provide quantified descriptions. Hypotheses accompanied by quantitative models serve as foundations for establishing models suitable for clinical applications.

Toward clinical applications, it is crucial for researchers and clinicians to establish a model that describes the relationship between radiobiological effects and radiation features. For instance, models have been developed to calculate the relative biological effectiveness (RBE) of proton and heavy ion radiotherapy, enabling the prediction of radiobiological effects based on microdosimetric parameters \citep{Elsasser_2007, Hawkins_1996}. Similarly, the application of FLASH radiotherapy necessitates models that elucidate how irradiation features, such as total dose, dose rate, and irradiation time, as well as other chemical or biological-related factors, determine the FLASH effect. 
In pursuit of establishing a practical model for clinical application, researchers have attempted to quantitatively predict the clinical effect using experimental data, simulations, and radiobiological models. 
The FLASH modifying factor was introduced and calculated based on the summary of experimental data \citep{bohlen_normal_2022}. 
The tumor control probability of FLASH effect was analyzed based on a model named as UNIVERSE \citep{Liew_2023}. 
A formalism \citep{bohlen_minimal_2022} was developed that quantifies the minimal normal tissue sparing of the FLASH effect required to compensate for hypofractionation. 
However, their models and analyses fall short of linking the FLASH effect to theoretical considerations, thus failing to capture the influence of mechanism factors on FLASH effect.

In this study, we quantitatively analyze the mathematical models of FLASH effect based on the oxygen depletion hypothesis and the radical recombination-antioxidants hypothesis, and subsequently develop the corresponding clinical models to describe the impact of FLASH irradiation features and mechanism factors on the FLASH effect. Based on the models and mathematical analysis, we propose the implementation of hypotheses for clinical application, including the criteria for data filtration, suggestions for systematic experiments and also the pipelines for parameter fitting and FLASH effect prediction.

\section{Materials and Methods}
\indent The actual scenario of FLASH irradiation is inherently complex. To conduct a mathematical analysis of the FLASH effect using theoretical hypotheses, it is necessary to simplify the intricate situation into concise mathematical representations. In this study, we abstract the complex models of the oxygen depletion hypothesis and radical recombination-antioxidants hypothesis into concise equations. Then we solve these equations to examine the impact of radiation features and other factors on the FLASH effect. 
The damage reduction factor (DRF) is defined to quantitatively evaluate the FLASH effect. The formulas of DRF vs. radiation features and other factors were derived.  
Moreover, we show how to implement the mechanism hypothesis-based models into clinical applications through a systematic pipeline to acquire experimental data.

\subsection{Damage reduction factor}
We introduce the damage reduction factor (DRF) in \Eref{DMF_DEF} to quantify the FLASH effect for specific dose and exposure times. The two hypotheses primarily focus on the differences in damage caused by FLASH versus CONV irradiation. The DRF is defined as the ratio of damage from FLASH irradiation to that from CONV irradiation for an equivalent dose, as depicted in \Eref{DMF_DEF}, where $T$ refers to the exposure time of FLASH irradiation.

\begin{equation}
\label{DMF_DEF}
DRF=\frac{Damage_{FLASH}(D,T)}{Damage_{CONV}(D)}
\end{equation}

It should be noted that the two hypotheses do not offer an exact definition of “damage”. The term “damage” used here is an abstract variable that can represent any form of damage that increases roughly linearly with dose, such as the number of double-strand breaks, or the concentrations of oxidized nucleic acids, proteins, and lipids. 
This definition maintains the additivity of damage, as the total biological effect during the entire irradiation period must be taken into account.
 
Furthermore, we derive the limit of \Eref{DMF_DEF} for $T\to 0$, as detailed in Equation \Eref{DMF_MIN}, to quantify the maximal change in biological effect that can be induced by a given dose of FLASH irradiation.

\begin{equation}
\label{DMF_MIN}
DRF_{min}=\lim_{T\to 0}\frac{Damage_{FLASH}(D,T)}{Damage_{CONV}(D)}
\end{equation}

\subsection{Model based on oxygen depletion hypothesis}

\subsubsection{Mathematical model}
\ 
\newline
\indent The oxygen depletion hypothesis is a widely discussed explanation for the mechanism underlying the FLASH effect. It suggests that the swift administration of radiation in FLASH radiotherapy leads to a substantial reduction in tissue oxygen levels, primarily because of the intense radiation-induced oxygen consumption. Due to the limited speed of oxygen diffusion, the oxygen supply cannot be replenished quickly enough in the irradiated area \citep{ZhuH_2021, Favaudon_2021}. Assuming immediate hypoxia in cells subjected to FLASH radiation, this hypothesis proposes that these cells demonstrate enhanced radioresistance compared to those exposed to CONV radiation. The two processes, radiation-induced oxygen consumption and subsequent oxygen recovery, are central to this hypothesis.
To establish a quantitative model for analysis and clinical application, 
we encapsulate the key oxygen dynamics at a tissue point into \Eref{ROD1}, in line with the oxygen depletion hypothesis. 
This equation directly integrates findings from numerical solutions to equations describing oxygen diffusion, metabolic consumption, and reactions \citep{hu_computational_2022}. Although oxygen recovery is inherently a spatial process, the equation here does not include the spatial term because it is included implicitly in terms of oxygen recovery and initial oxygen concentration.
  
\begin{equation}
\label{ROD1}
\frac{\mathrm{d}p(t)}{\mathrm{d}t}=Recovery(t, p)-ROC(t, p)
\end{equation}
where $p(t)$ is the concentration of oxygen at the time point $t$; $Recovery(t, p)$ is the term related to oxygen recovery in the tissue; $ROC(t, p)$ is the radiolytic oxygen consumption rate.  
The differential equation has an initial condition as listed in \Eref{ROD_IC}.
\begin{equation}
\label{ROD_IC}
p(t=0)=p_0
\end{equation}
where $p_0$ represents the initial oxygen concentration of the tissue before irradiation, which is influenced by factors such as the oxygen concentration in the capillaries, the density of capillaries, the rate of oxygen diffusion, the rate of metabolic oxygen consumption, and the distance from the tissue point to the nearest capillary.

Previous studies indicate that the process of oxygen recovery is mainly attributed to oxygen diffusion. 
The temporal profile of oxygen concentration approximately follows an exponential decay pattern \citep{hu_computational_2022, ZhuH_2021, Pratx_2019}. 
The term of oxygen recovery in \Eref{ROD1} can be described by \Eref{ROD_dif}.
\begin{equation}
\label{ROD_dif}
Recovery(t, p)=\lambda [p_0 - p(t)]
\end{equation}
where $\lambda$ is the characteristic constant for oxygen diffusion, determined by factors such as the density of microvessels in the tissue, the diffusivity of the medium and the tortuosity of cells. 
The parameter $\lambda$ encapsulates the spatial aspects of oxygen concentration variations in the model in an implicit manner.

The radiolytic oxygen consumption is mainly attributed to the reactions between radiation-induced radicals and oxygen \citep{Wardman_2020,Wardman_2021,Wardman_2022}. In many related studies, the radiolytic oxygen consumption rate ($\mu$M/Gy) is often treated as a constant \citep{Pratx_2019, ZhuH_2021}. However, this assumption is invalid in the cells where the oxygen concentration is low. Some studies set the oxygen consumption rate proportional to the oxygen consumption \citep{Petersson_2020}, which also cannot reflect the real condition when the oxygen concentration is high. 
Experiments have observed different radiolytic oxygen consumption rates under different initial oxygen concentrations\citep{el_khatib_direct_2023,jansen_does_2021, slyke_oxygen_2022}. The difference can be attributed to two effects: reactions between radiation-induced radicals and competition between oxygen and other cellular compounds (for example, antioxidants) reacting with radicals\citep{Wardman_2016}. The short lifetime of radicals (e.g., ·OH, ·H, $e_{aq}$) resulting from water radiolysis, typically in the nanosecond range \citep{roots_estimation_1975}, does not align with the microsecond timescale of radiolytic oxygen consumption. Reactions among these radicals cannot play the primary role in the process. Thus, it is derived that the competition between oxygen and cellular compounds is the main factor influencing radiolytic oxygen consumption. The amount of radicals is proportional to the radiation dose. 

Based on the above analysis, we use a fractional formula to describe the competition between oxygen and other cellular compounds as the formula used by Zou et al.\citep{Zou_2022}, which is the solution to the equation describing the competition between oxygen and antioxidants to react with radiation-induced radicals. This formula also corresponds to the results of radiolytic oxygen consumption vs. dose \citep{jansen_does_2021, slyke_oxygen_2022}. The term of radiolytic oxygen consumption can be described by \Eref{ROD_ROD}.
\begin{equation}
\label{ROD_ROD}
ROC(t, p)=g_1\dot{D}(t)\cdot\frac{p(t)}{p(t)+A}
\end{equation}
where $g_1$ is the yield of radiolytic oxygen consumption, which is related to the radiation-induced radicals; $\dot{D}(t)$ is the dose rate at the time point $t$; $A$ represents the equivalent concentration of other cellular compounds competing with oxygen to react with radiation-induced radicals.

Previous studies have shown that the average dose rate during irradiation plays a crucial role in the FLASH effect \citep{hu_computational_2022, Karsch_2022}. In this study, we simplify the equation by assuming a constant dose rate throughout the irradiation, represented by \Eref{ROD_DR}. 
This simplification proved valid for beams with much shorter pulse intervals than characteristic  time of oxygen recovery ranging from hundreds of milliseconds to several seconds \citep{hu_computational_2022}. The requirement of the irradiation time structure well applies to the two types of beams commonly used in the FLASH experiments currently, i.e., linac electron beams with the pulse interval of $\sim$5 ms, and cyclotron protons with continuous wave mode.
On the other hand, if the pulse interval is long enough, the whole therapy can be treated as the combination of several independent FLASH irradiations. 
It should be noted that if the pulse interval is neither too long nor too short, or the pulse interval is short but the pulse intensity is also changing, the analytical method is not suitable and numerical solutions are preferable.

\begin{equation}
\label{ROD_DR}
\dot{D}(t)=\frac{D}{T}
\end{equation}
where $D$ is the total dose and $T$ is the total irradiation time.

With the above assumptions and simplifications, \Eref{ROD1} is converted into \Eref{ROD2}.
\begin{equation}
\label{ROD2}
\frac{\mathrm{d}p(t)}{\mathrm{d}t}=\lambda [p_0 - p(t)]-g_1\cdot\frac{D}{T}\cdot\frac{p(t)}{p(t)+A}
\end{equation}
 
\subsubsection{Radiobiological effects of FLASH and CONV irradiation}
\ 
\newline
\indent The radiobiological effect is often estimated based on the classical Alper's formula \citep{Alper_1956,Alper_1983} of the radiation oxygen effect listed in \Eref{ROD_Alper}. To keep the definition of $DRF$ and the additivity for evaluation of the overall effect of the whole irradiation, we use the $OER$ value in \Eref{ROD_Alper} to represent the damage.
\begin{equation}
\label{ROD_Alper}
OER = \frac{K+mp(t)}{K+p(t)}
\end{equation}
where $OER$ is the oxygen enhancement ratio, which represents the ratio of the damage under $p(t)$ oxygen to the damage in the hypoxic condition with the same dose delivered; $m$ is the maximal $OER$ and $K$ is the oxygen concentration at the half-maximal $OER$.

The relative amount of damage induced by the FLASH irradiation is calculated by the dose-averaged $OER$ over the whole irradiation period. For a constant dose rate scenario, damage can be calculated by \Eref{ROD_BE}. 
\begin{equation}
\label{ROD_BE}
Damage=\frac{1}{T} \int_0^T\frac{K+mp(t)}{K+p(t)}\mathrm{d}t
\end{equation}

As for the CONV condition, because of the low dose rate delivery, the oxygen concentration remains almost unchanged during the irradiation. Thus, the biological effect can be represented as \Eref{ROD_CONV},
\begin{equation}
    \label{ROD_CONV}
    Damage_{CONV}=\frac{K+mp_0}{K+p_0}
\end{equation} 

To estimate the maximal FLASH effect, we calculate the limit of the function defined by \Eref{ROD_BE} by setting $T\to 0$. 

Within the extremely short irradiation duration, the oxygen recovery can be ignored. The relationship between oxygen concentration at the end of irradiation and the total dose can be derived by solving \Eref{ROD3} with initial condition \Eref{ROD_IC}.
\begin{equation}
    \label{ROD3}
    \frac{\mathrm{d}p(t)}{\mathrm{d}t} = -\frac{g_1D}{T}\cdot\frac{p(t)}{p(t)+A}
\end{equation}

By setting $t=T$ in the solution of \Eref{ROD3}, we can calculate the oxygen concentration at the end of the irradiation, $p_T$, to estimate the maximal change of oxygen concentration during FLASH irradiation.

\subsection{Model based on radical recombination-antioxidants hypothesis}
\subsubsection{Mathematical model}
\ 
\newline
\indent The radical recombination-antioxidants hypothesis explains the protective effect of normal tissue by the recombination of peroxyl radicals (including superoxide anion)\citep{Labarbe_2022}. Additionally, it elucidates the loss of this protective effect in tumors due to high levels of antioxidants present \citep{hu_radical_2023}. 
Antioxidants act as a scavenger of peroxyl radicals. They determine the average lifetime of peroxyl radicals and the portion of radical recombination reactions.
The main point of establishing the mathematical model is to appropriately describe the reaction of peroxyl radicals. Despite the inherent complexity of these reactions, we can simplify the reaction model into three primary processes: reaction of peroxyl radical with antioxidant, peroxyl radical recombination and generation of peroxyl radical by irradiation, as demonstrated in a concise form in \Eref{RA1}. The lifetimes of radicals from water radiolysis are too short to alter the reactions of peroxyl radicals, they are not included in the model.
\begin{equation}
    \label{RA1}
    \frac{\mathrm{d}R(t)}{\mathrm{d}t} = -k_1R(t)-k_2[R(t)]^2+g_2\dot{D}(t)
\end{equation}
where $R(t)$ is the concentration of peroxyl radicals; $g_2$ is the yield of peroxyl radicals;  because the concentration of radiation-induced peroxyl radical ($\sim\mu$M) \citep{isildar_oxygen-uptake_1982} is much lower than the concentration of cellular antioxidants ($\sim$mM) \citep{DING2021128880}, the concentration of the antioxidant is treated as a constant, and thus here we use the first-order rate constant of peroxyl radicals reacting with antioxidants (including the effects of concentration and rate constant), $k_1$ (s$^{-1}$), to describe the reaction; $k_2$ (M$^{-1}$s$^{-1}$)is the second-order rate constant of peroxyl radical recombination, which roughly sums the reactions of different types of reactants;  $g_2$ is the yield of peroxyl radicals per unit dose with the assumption that the concentration of radical is proportional to the dose. Noted that the radical recombination is modeled by the $k_2[R(t)]^2$ according to the reaction rate equation \citep{Labarbe_2022,Wardman_2022}.

According to the same reason described in the model of oxygen depletion, we consider the radiation delivered at a constant dose rate. Thus, \Eref{RA1} can be divided into two stages: \Eref{RA_2_1} during irradiation and \Eref{RA_2_2} after irradiation.
\begin{equation}
    \label{RA_2_1}
    \frac{\mathrm{d}R(t)}{\mathrm{d}t} = -k_1R(t)-k_2[R(t)]^2+\frac{g_2D}{T},\\
    R(0)=0
\end{equation}
\begin{equation}
    \label{RA_2_2}
    \frac{\mathrm{d}R(t)}{\mathrm{d}t} = -k_1R(t)-k_2[R(t)]^2,\\
    R(T)=R_T
\end{equation}
where $R_T$ is the concentration of peroxyl radicals at the end of irradiation, which can be calculated by the solution of \Eref{RA_2_1}. 

\subsubsection{Radiobiological effect of FLASH and CONV irradiation}
\ 
\newline
\indent The radiation-induced damage can be classified into two types: the peroxyl radical-dependent damage and the peroxyl radical-independent damage. Peroxyl radical-dependent damage can be estimated by the concentration-time integral of peroxyl radical concentration defined in\Eref{RA_AUC} as Labarbe et al. \citep{Labarbe_2022}. 
\begin{equation}
    \label{RA_AUC} 
    AUC[ROO\cdot] = \int_0^{+\infty}R(t)\mathrm{d}t
\end{equation}
where $AUC[ROO\cdot]$ is considered to be proportional to peroxyl radical-dependent damage, e.g. the amount of oxidative damaged nucleic acids, proteins and lipids \citep{hu_radical_2023}.
Then the total damage can be calculated as \Eref{RA_DAM}.
\begin{equation}
    \label{RA_DAM}
    Damage = f_1\cdot AUC[ROO\cdot] + f_2D
\end{equation}
where $f_1$ is the factor related to radical-dependent damage; $f_2$ is the factor related to radical-independent damage. The relative values of $f_1$ and $f_2$ are mainly determined by the radiation quality, e.g., the energy and type of the particle. Using this formula, we can calculate the radiobiological effect of FLASH irradiation by a given dose and time.

For CONV irradiation, we calculate the limit of the damage function defined by \Eref{RA_DAM} to estimate the radiobiological effect of CONV irradiation.
\begin{equation} \label{RA_L1}
\lim_{T\to +\infty}Damage 
\end{equation}

\subsection{Parameters of the models}
The primary objective of this study is to derive mathematical formulas that quantify the relationships between key factors in the hypotheses and the FLASH effect. The derived formulas analytically describe these relationships. In order to provide a visual representation of the formulaic trends, figures are utilized. It is important to note that these figures are designed to illustrate the shape of the curves described by the formulas, rather than to reflect real-world scenarios. Generating these figures necessitates the explicit values of parameters. While some of the parameters are determined based on estimates derived from existing literature, others are assigned values without specific justifications, as these figures solely serve to depict the curve shapes within the formulas. The parameters employed in the model of the oxygen depletion hypothesis are listed in \tref{table:PROD}, while the parameters utilized in the model of the radical recombination-antioxidants hypothesis are listed in \tref{table:PRA}.  

\begin{table}
\caption{Parameters in the model of the oxygen depletion hypothesis}
\label{table:PROD}
\begin{tabular*}{\textwidth}{c p{7.5cm} p{5cm}}
\br
Parameter &  Description &Value\\
\mr
$\lambda$ &  feature constant of oxygen recovery & 7.9 s$^{-1}$ \citep{Petersson_2020}\\
$g_1$ & yield of radiolytic oxygen consumption & 0.3 $\mu$M/Gy (0.22 mmHg/Gy) \citep{Cao_2021}\\
$A$ & equivalent concentration of other cellular compounds competing with oxygen& 5 $\mu$M $^{*}$ (3 mmHg)\citep{Prise_1992}\\ 
$p_0$ & initial oxygen concentration of tissue& 10 $\mu$M $^{**}$(7 mmHg)\\
$D$ & total dose& 20 Gy $^{***}$\\
$K$ & oxygen concentration at the half-maximal $OER$& 7.2 $\mu$M (5 mmHg)\citep{Alper_1956}\\
$m$ & maximal $OER$& 2.9 \citep{Alper_1956}\\
\br
\end{tabular*}
\begin{tablenotes}
\item[]$^*$roughly estimated by the reaction between DNA radical and GSH; $^{**}$set as an example; $^{***}$set as an example.
\end{tablenotes}
\end{table}

\begin{table}
\caption{Parameters in the model of the radical recombination-antioxidants hypothesis}
\label{table:PRA}
\begin{tabular*}{\textwidth}{c p{7.5cm} p{5cm}}
\br
Parameter &  Description & Value\\
\mr
$g_2$ &  yield of peroxyl radical & 0.3 $\mu$M/Gy \citep{Cao_2021}\\
$k_1$ & first-order rate constant of reaction between peroxyl radical and antioxidant & 8.5 s$^{-1}$ \citep{hu_radical_2023}\\
$k_2$ & second-order rate constant of radical recombination & $1.0\times 10^6 M^{-1}\cdot s^{-1}$ \citep{Neta_1990,Hasegawa_1978}\\
$D$ & total dose& 20 Gy\\
$f_1$ & factor related to radical-dependent damage&  1 $^*$\\
$f_2$ & factor related to radical-independent damage& 0.3 $\mu$M$\cdot$s/Gy $^{**}$\\
\br
\end{tabular*}
\begin{tablenotes}
\item[]$^{*, **}$set as examples to show the shapes of curves in figures. 
\end{tablenotes}
\end{table}

\subsection{Implementation of the hypotheses for clinical applications}
\indent Clinical applications of radiotherapy necessitate quantitative models to describe the dose-biological effectiveness relationship. These models rely on data obtained from experiments and clinical trials. 
In particular, forthcoming experiments and clinical trials involving FLASH radiotherapy are expected to yield datasets encompassing various total doses and dose rates. 
Based on these datasets, formulas can be used to quantitatively predict the FLASH effect for any given dose and irradiation time. 

In this work, we have developed concise models based on two main hypotheses. These models facilitate both the fitting of data points and the prediction of the FLASH effect under specific irradiation conditions, considering given total doses and dose rates. 
The derivation and analysis of these models indicate that the FLASH effect is determined by several key biological factors. The key factors among samples in one group data should remain consistent to avoid the influence attributed to the inconsistency of key factors. Predicting the FLASH effect based on the models also requires the consistency of these key factors. Our results provide criteria of data filtration for obtaining models' parameters and criteria to judge whether the model can be applicable to a specific patient.
These aspects are crucial for enhancing the accuracy and reliability of the model prediction.
 
Then, we provide suggestions for designing systematic experiments to obtain data points. 
According to the formulas derived, pipelines about how to fit the parameters of the models and utilize the models for the FLASH effect prediction are proposed. 
The three-step process, i.e., data filtration, systematic experiment design, and parameter fitting, indicates how to implement the mechanism hypotheses for clinical applications.

\section{Results and Discussions}
\indent In this section, we present the results of the derivation. The detailed derivation process is documented in the \textbf{Supplementary document}.

\subsection{Mathematical analysis of the model based on the oxygen depletion hypothesis}

\subsubsection{Solution of the equation}
\ 
\newline 
\indent The solution of \Eref{ROD2} is shown in \Eref{RROD1}. Obtaining an explicit solution is not attainable, so we list an implicit solution here.
\begin{equation}
\label{RROD1}
-\frac{A+p_1}{p_1-p_2}\log{|\frac{p(t)-p_1}{p_0-p_1}|}+\frac{A+p_2}{p_1-p_2}\log{|\frac{p(t)-p_2}{p_0-p_2}|}=\lambda t
\end{equation}
where $p_1$ and $p_2$ are two constants calculated by \Eref{RROD2_1} and \Eref{RROD2_2}.
\numparts
\begin{eqnarray}
\label{RROD2_1} 
p_1=\frac{- A T \lambda - g_1 D + T \lambda p_0 - \sqrt{4 A T^{2} \lambda^{2} p_{0} + \left(A T \lambda + g_1 D - T \lambda p_{0}\right)^{2}}}{2 T \lambda} \\
p_2=\frac{- A T \lambda - g_1 D + T \lambda p_0 + \sqrt{4 A T^{2} \lambda^{2} p_{0} + \left(A T \lambda + g_1 D - T \lambda p_{0}\right)^{2}}}{2 T \lambda} \label{RROD2_2}
\end{eqnarray}
\endnumparts

Then we calculated the integral in \Eref{ROD_BE} by changing the integration variable from $t$ to $p$ as shown in \Eref{RROD_int}.
\begin{equation}
    \label{RROD_int}
    Damage = m-\frac{(m-1)K}{T}\int_{p_0}^{p_T}\frac{1}{p+K}\cdot\frac{\mathrm{d}t}{\mathrm{d}p}\cdot \mathrm{d}p
\end{equation}
The result of the integral is shown in \Eref{RROD_intR}.
\begin{equation}
\eqalign{ Damage
&=m-(m-1)K[-\frac{A-K}{T\lambda(K+p_1)(K+p_2)}\log{\frac{K+p_T}{K+p_0}} \cr
&-\frac{A+p_1}{T\lambda(K+p_1)(p_1-p_2)}\log{|\frac{p_T-p_1}{p_0-p_1}|} \cr
&+ \frac{A+p_2}{T\lambda(K+p_2)(p_1-p_2)}\log{|\frac{p_T-p_2}{p_0-p_2}|}]} \label{RROD_intR}
\end{equation}

In the calculation with computer programs like MATLAB, we find that for the conditions of low total dose or long irradiation time, effective results cannot be obtained through the forms of \Eref{RROD1} and \eref{RROD_intR} due to the limited numerical precision of the computer. 
Thus, we use some tricks to conduct the calculation as follows.

For low total dose, we use the first-order Taylor expansion of \Eref{RROD1} to calculate $p(t)$, which is shown in \Eref{RROD_APP}.
\begin{equation}
    \label{RROD_APP}
    p(t)\approx -\frac{g_1 D p_0}{A + p_0}\cdot \frac{t}{T}+p_0
\end{equation}

With a similar mathematical trick with \Eref{RROD_APP}, the biological effect of the above-mentioned condition ($p_T\to p_0$) can be calculated by \Eref{RROD_APPBE}.

\begin{equation}
    \label{RROD_APPBE}
    \eqalign{ Damage\approx m - \frac{K(A+p_0)(m-1)}{g_1Dp_0}
    \log{[1-\frac{g_1 D p_0}{(A+p_0)(K+p_0)}]} }
\end{equation}

For long irradiation time, the oxygen recovery and radiolytic oxygen consumption reach equilibration, i.e., $\lambda(p_T-p_0)\approx g_1Dp_T/[T(A+p_T)]$. The term, $p_T$, can be calculated by \Eref{RROD_APP2}.

\begin{equation}
    \label{RROD_APP2}
    p_T\approx p_2
\end{equation}

The damage can be calculated by \Eref{RROD_APPBE2} according to \Eref{RROD_APP2}, \Eref{RROD_intR} and \Eref{RROD1}.

\begin{equation}
    \label{RROD_APPBE2}
    \eqalign{ Damage\approx m-(m-1)K[-\frac{A-K}{T\lambda(K+p_1)(K+p_2)}\log{\frac{K+p_2}{K+p_0}} \cr
-\frac{A+p_1}{T\lambda(K+p_1)(p_1-p_2)}\log{|\frac{p_2-p_1}{p_0-p_1}|} \cr
+\frac{1}{K+p_2}+ \frac{A+p_1}{T\lambda(K+p_2)(p_1-p_2)}\log{|\frac{p_2-p_1}{p_0-p_1}|}]}
\end{equation}

\subsubsection{Radiolytic oxygen consumption}
\
\newline
Radiolytic oxygen consumption is one of the key factors in the oxygen depletion hypothesis. By setting $t=T$ in the \Eref{RROD1}, we can calculate the radiolytic oxygen consumption (net change of oxygen concentration during the irradiation, which includes the effect of oxygen recovery) induced by a given dose ($D$) and irradiation time ($T$). Here, we define $\Delta p=p_0-p_T$ to represent the radiolytic oxygen consumption. The maximal radiolytic oxygen consumption, $\Delta p_{max}$, is calculated by taking the limit for $T\to 0$. 

\begin{figure} 
    \centering
    \includegraphics[scale=0.4]{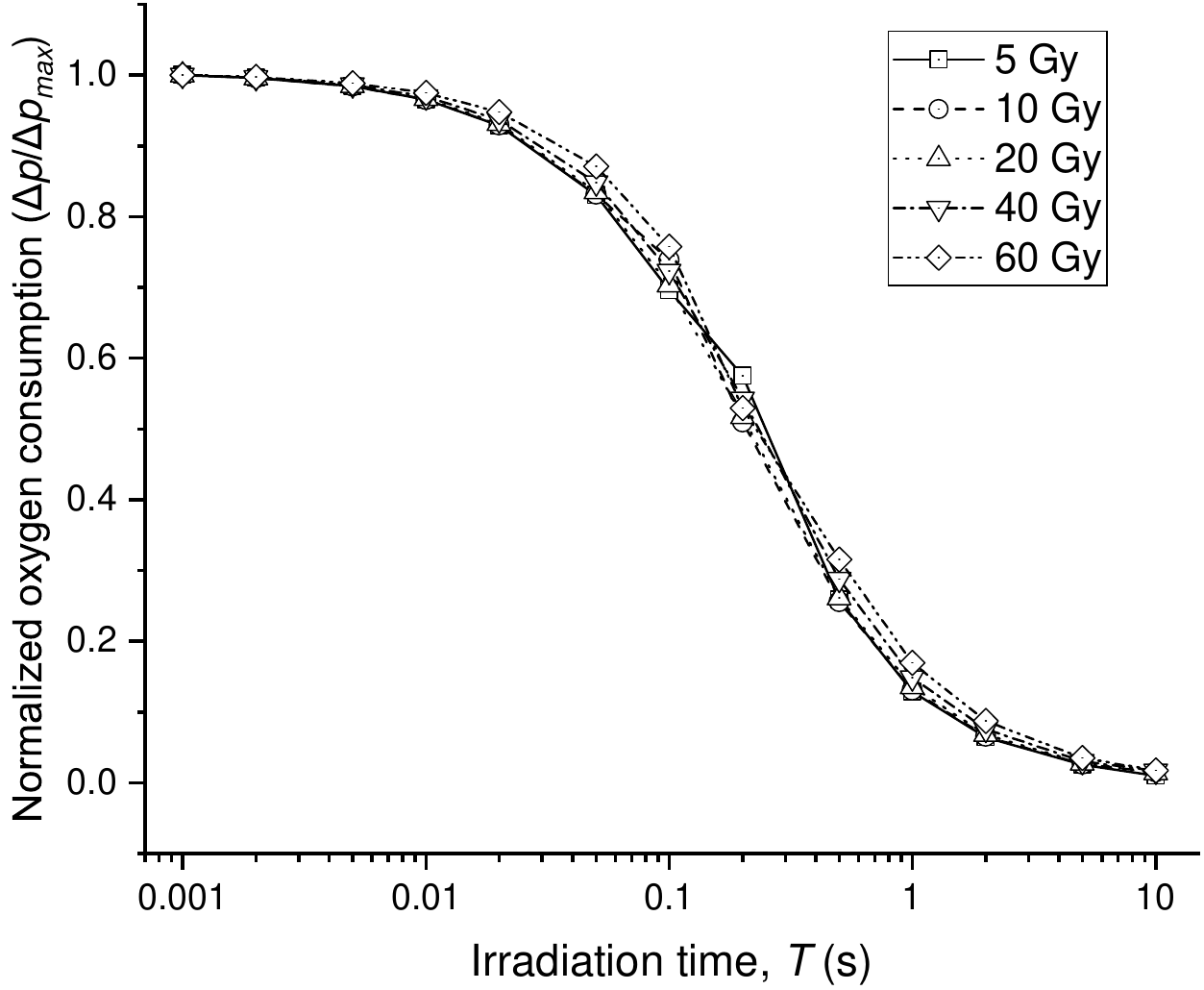}   
    \caption{
    Relative radiolytic oxygen consumption as a function of irradiation time (0.01 to 10 s) for radiation doses ranging from 5 to 60 Gy, with an initial oxygen concentration ($p_0$) as 10 $\mu$M (7 mmHg) }
    \label{fig:PTROD}
\end{figure}

First, to show the impact of irradiation time on the radiolytic oxygen consumption, 
by setting the initial oxygen concentration ($p_0$) as 10 $\mu$M (7 mmHg),
we calculate a series of radiolytic oxygen consumption ($\Delta p$) for a range of irradiation times spanning milliseconds to several seconds, under total doses ranging from 5 Gy to 60 Gy.
The resulting curves of normalized oxygen consumption, i.e., $\Delta p/\Delta p_{max}$, are plotted against irradiation time for different dose deliveries in \fref{fig:PTROD}.

The results of the mechanism hypothesis-based calculations highly reflect the core principle of the oxygen depletion hypothesis. 
The curves exhibit an identical inverse “S” shape, indicating that shorter irradiation time results in greater oxygen consumption due to the limited rate of oxygen recovery. 
Additionally, the curves are nearly overlapping across different doses. This suggests that the temporal characteristic of radiolytic oxygen consumption is largely dose-independent. The oxygen recovery process determines how the irradiation time impacts the radiolytic oxygen consumption, which is almost independent of dose.

Furthermore, we study the influences of initial oxygen concentration ($p_0$) and dose ($D$) on the maximal oxygen consumption, $\Delta p_{max}$. The implicit solution \Eref{ROD3} is listed in \Eref{ROD_dp}.

\begin{equation}
    \label{ROD_dp}
    \Delta p_{max}-A\log(1-\frac{\Delta p_{max}}{p_0})=g_1D  
\end{equation}

\begin{figure}
    \centering
    \includegraphics[scale=0.4]{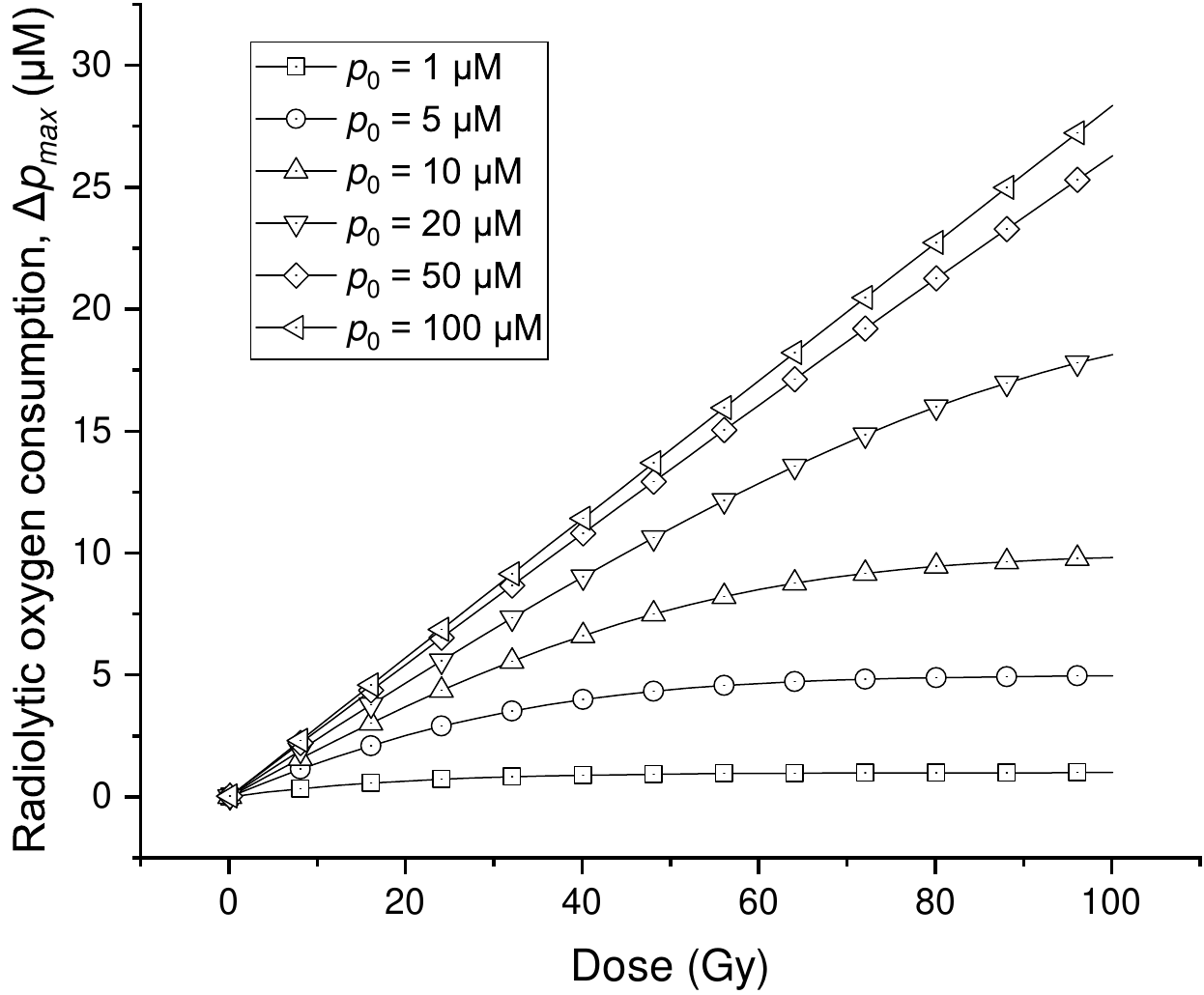}
    \caption{Maximal radiolytic oxygen consumption (irradiation time $T\to 0$) induced by FLASH irradiation with a series of total doses under different initial oxygen concentrations, $p_0$,  from 1 $\mu$M to 100 $\mu$M ($A = 5$ $\mu$M)}
    \label{fig:ROC}
\end{figure}

With different initial oxygen concentrations ranging from 1 to 100 $\mu$M (0.7-70 mmHg), the values of $\Delta p_{max}$ against total doses from 0 to 100 Gy are calculated based on the \Eref{ROD_dp}.
As shown in \fref{fig:ROC},
the resulting curves indicate that $\Delta p_{max}$ is no longer proportional to the dose magnitude when the dose has already exceeded the amount required to make the cell hypoxic. 
The saturation effect occurs significantly for low values of initial oxygen concentration. For the simulation parameters used in this study, it can be seen in the curves with the initial oxygen concentration lower than 20 $\mu$M (14 mmHg).

Additionally, to show the influence of the competing reactions for the radiolytic oxygen reaction, we change the equivalent concentration of other cellular compounds ($A$) from 1.0 to 20.0 $\mu$M (0.7-14 mmHg) and calculate the corresponding $\Delta p_{max}$ induced by 20 Gy FLASH irradiation under different initial oxygen concentrations. 

\begin{figure}
    \centering
    \includegraphics[scale=0.4]{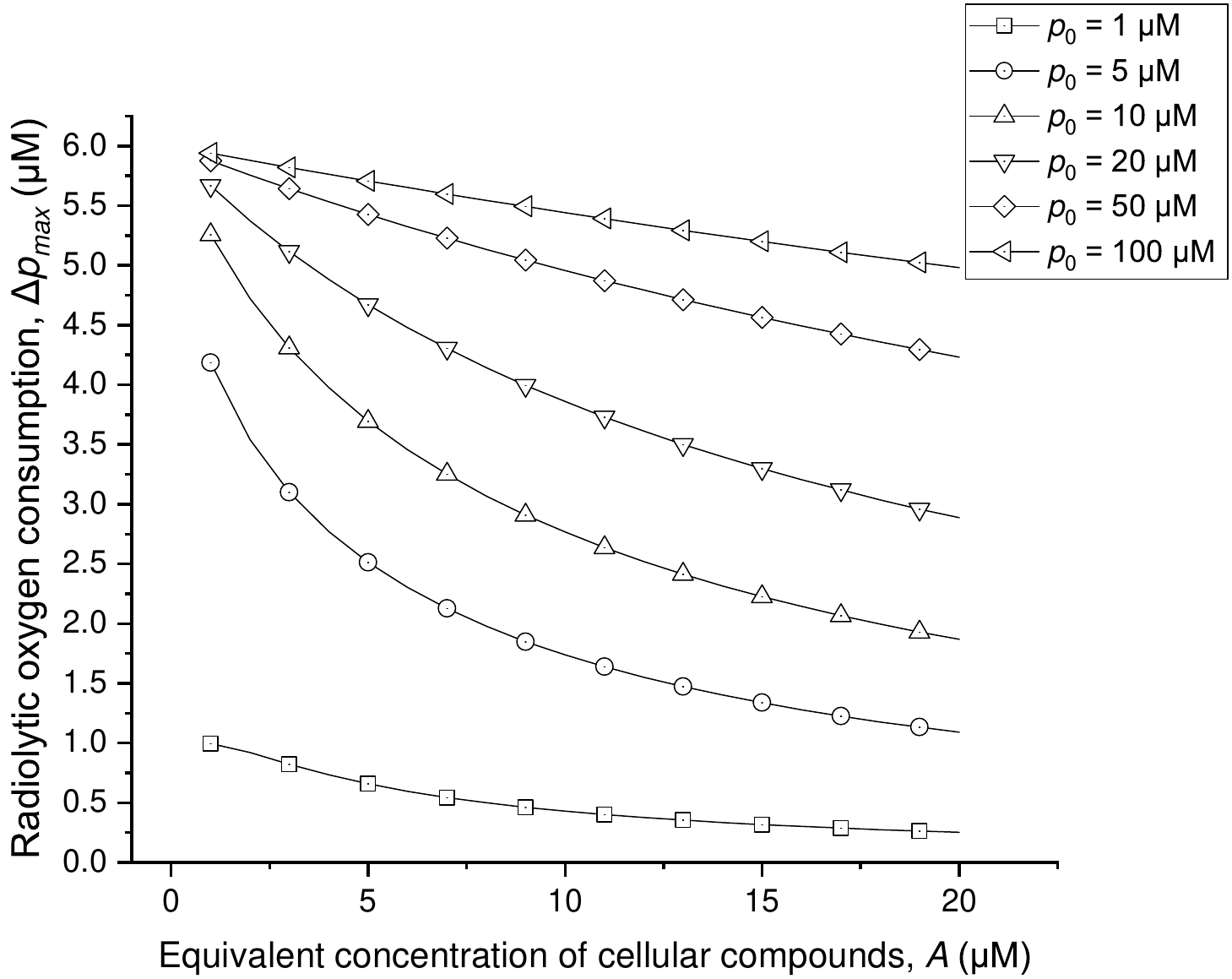}
    \caption{Maximal radiolytic oxygen consumption induced by 20 Gy FLASH irradiation in the cells with the equivalent concentration of cellular compounds, $A$, ranging from 1.0 to 20.0 $\mu$M (0.7-14 mmHg) under different initial oxygen concentrations}
    \label{fig:ROC_A}
\end{figure}

As shown in \fref{fig:ROC_A}, the competition between oxygen and other cellular compounds influences radiolytic oxygen consumption greatly when the initial oxygen concentration is low ($<20 
\mu$M for the parameters in this study). 
The radiolytic oxygen consumption cannot be regarded as a constant for different conditions of the cellular composition. 

\subsubsection{Radiobiological effects of FLASH and CONV irradiation}
\ 
\newline

\begin{figure}
    \centering
    \includegraphics[scale=0.4]{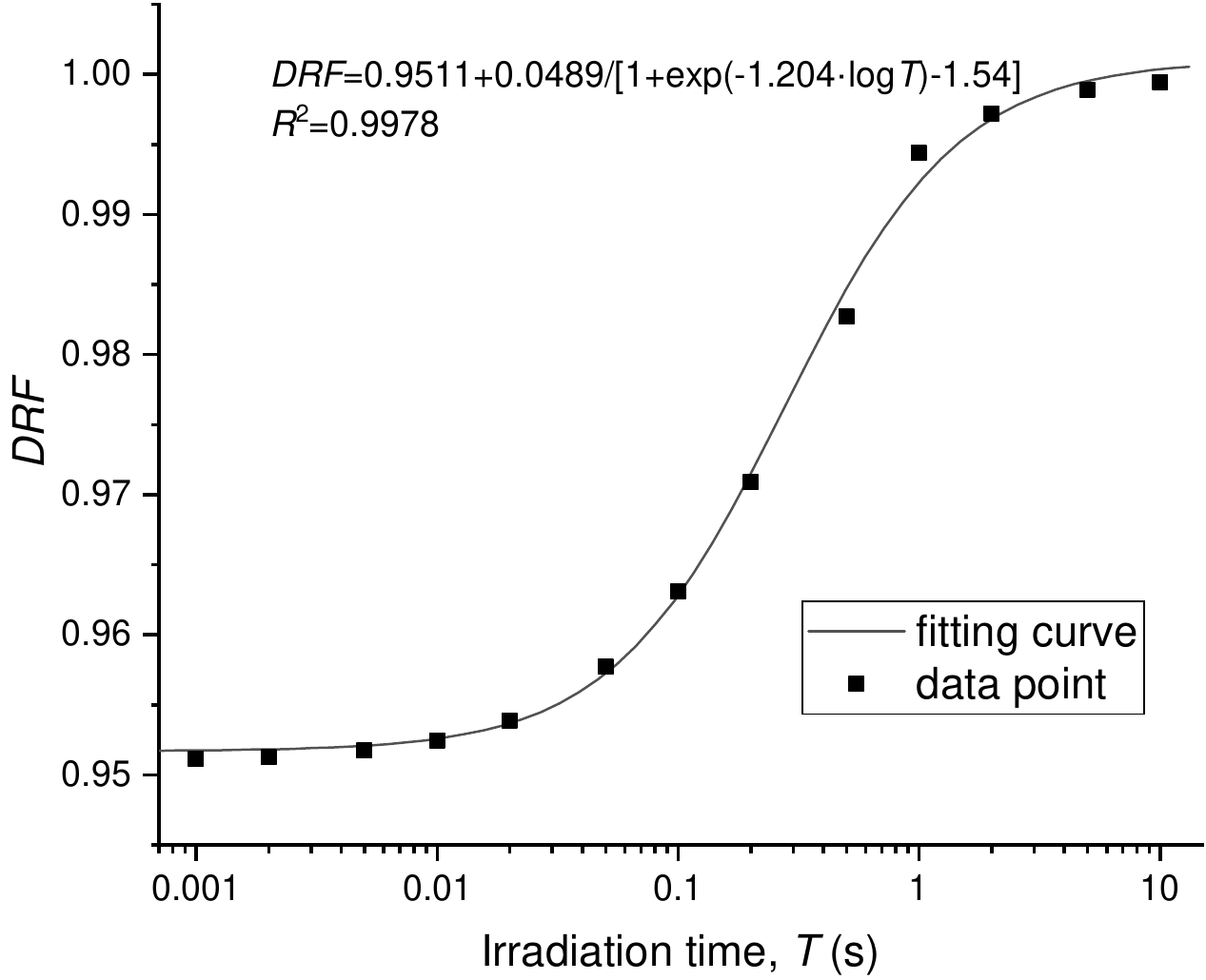}
    \caption{$DRF$ of 20 Gy FLASH beam with irradiation time ranging from 0.001 to 10 s and the fitting curve of the data points, when initial oxygen concentration, $p_0$, is 10 $\mu$M (7 mmHg).}
    \label{fig:DMF_T}
\end{figure}

The impact of irradiation time ($T$) on the FLASH effect ($DRF$) can be derived as \Eref{DRF_ROD} by \Eref{RROD_intR} (damage of FLASH) and \Eref{ROD_CONV} (damage of CONV). Data points of $DRF$s vs. $T$ for an example dose setup are shown in \fref{fig:DMF_T}. The total dose is set to 20 Gy, and the initial oxygen concentration is set as $p_0=10\mu$M.

\begin{equation}
    \label{DRF_ROD}
    \eqalign{ 
    DRF(T) 
    &= \frac{(K+p_0)m}{K+mp_0} -
    \frac{(K+p_0)(m-1)K}{K+mp_0}[-\frac{A-K}{T\lambda(K+p_1)(K+p_2)}\log{\frac{K+p_T}{K+p_0}} \cr
    &-\frac{A+p_1}{T\lambda(K+p_1)(p_1-p_2)}\log{|\frac{p_T-p_1}{p_0-p_1}|} \cr
    &+ \frac{A+p_2}{T\lambda(K+p_2)(p_1-p_2)}\log{|\frac{p_T-p_2}{p_0-p_2}|}]
    } 
\end{equation}

The relationship of $DRF$ vs. $T$ can be approximated by a relatively simple formula (\Eref{RROD_FITBE}). The fitting curve of this formula is plotted in \fref{fig:DMF_T} with fitting parameters listed in \tref{table:fit_ROD}.

 \begin{equation}
     \label{RROD_FITBE} 
     DRF(T) = DRF_{min}+\frac{1-DRF_{min}}{1+\exp (\xi_1\log T+\xi_2)}
 \end{equation}
 where $DRF_{min}$ is the minimal $DRF$ for a given dose with extremely high dose rate irradiation, i.e., $T \to 0$; 
 $\xi_1$ and $\xi_2$ are dose-independent constants, corresponding to the dose-independence of the time characteristic of radiolytic oxygen consumption shown in \fref{fig:PTROD}.

\begin{table}
\caption{Parameters and goodness of fitting curves of $DRF$ vs. $T$}
\label{table:fit_ROD}
\begin{tabular*}{\textwidth}{@{}l*{15}{@{\extracolsep{0pt plus12pt}}l}}
\br
Parameter & Value\\
\mr
$DRF_{min}$ & 0.9511\\
$\xi_1$ & -1.204\\
$\xi_2$ & -1.54\\
$R^2$  & 0.9978\\
\br
\end{tabular*}
\end{table}

As for the choice of initial oxygen concentration, $10 \mu$M, in this example, though it only represents the situation for hypoxic tissue and tumors \citep{vaupel_oxygenation_1991}, we present the resulting curve because the change of biological effect is numerically larger and the trend can be observed more obviously than that for normal tissue situation. The average oxygen concentration of normal tissue is  $\sim 70 \mu$M (50 mmHg)  \citep{vaupel_oxygenation_1991}, under which the change of biological effect is too slight to show in the figure.

Then, the minimal damage of FLASH irradiation by a given dose is estimated by calculating the limit of \Eref{RROD_intR} when $T \to 0$. The result is shown in \Eref{RROD_FLASH}
\begin{equation}
    \label{RROD_FLASH}
    Damage_{min}=m-(m-1)[\frac{A-K}{g_1D}\log(1-\frac{\Delta p_{max}}{K+p_0})+\frac{g_1D-\Delta p_{max}}{g_1D}]
\end{equation}

The $DRF_{min}$ is given with \Eref{RROD_FLASH} deciding the $Damage$ induced by the same dose of CONV irradiation as shown in \Eref{RROD_DMF}. 
\begin{equation}
    \label{RROD_DMF}
    DRF_{min} = \frac{\left(K + p_{0}\right) \left(-(\frac{A-K}{g_1D}\log(1-\frac{\Delta p_{max}}{K+p_0})+\frac{g_1D-\Delta p_{max}}{g_1D}) \left(m - 1\right) + m\right)}{K + m p_{0}}
\end{equation}

\begin{figure}
 \centering
    \includegraphics[scale=0.4]{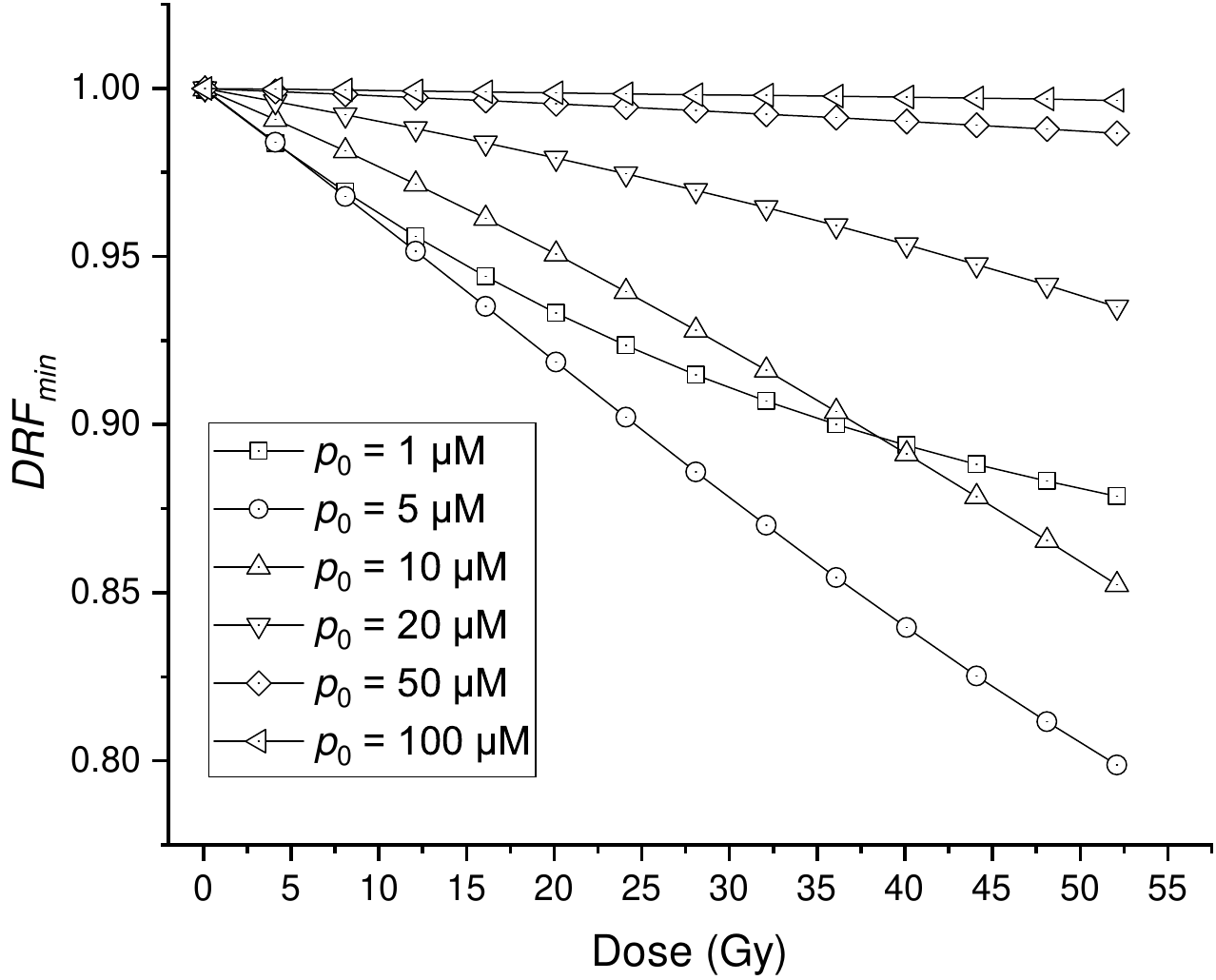}
    \caption{Minimal damage reduction ratio (irradiation time $T\to 0$), $DRF_{min}$, for different doses under initial oxygen concentration ranging from 1 to 100 $\mu$M (0.7-70 mmHg), predicted by oxygen depletion hypothesis}
    \label{fig:DMFROD}
\end{figure}

The $DRF_{min}$ for different doses and initial oxygen concentrations are shown in \fref{fig:DMFROD}. 
The curves indicate that the FLASH effect predicted by the oxygen depletion hypothesis is influenced by the initial concentration greatly. 
For normal oxygenated tissues with the median oxygen concentration of $70 \mu$M (50 mmHg) \citep{vaupel_oxygenation_1991}, the biological effect changes slightly. 
However, in hypoxic tissues or hypoxic parts of tumors where the typical oxygen concentration ranges from 0 to $10 \mu$M (7 mmHg)\citep{vaupel_oxygenation_1991, menon_integrated_2003}, the biological effect alters greatly, which contradicts the almost unchanged tumor control shown in FLASH effect experiments. Our results indicate that the oxygen depletion hypothesis cannot feasibly explain the FLASH effect that spares the normal tissue but keeps the equivalent tumor control.

\subsection{Mathematical analysis based on radical recombination-antioxidants hypothesis}
\subsubsection{Solution of the equation}
\ 
\newline
\indent
The solutions of the model based on the radical recombination-antioxidants hypothesis contain two parts. The first part is the solution of \Eref{RA_2_1}, which describes the reactions during irradiation ($0\leq t < T$). The solution is shown in \Eref{RRA_2_1}.
\begin{equation}
    \label{RRA_2_1}
     R(t)=\frac{\sqrt{4g_2Dk_2/T+k_1^2}/k_2}{C_1\exp(t\sqrt{k_1^2+4k_2g_2D/T})-1}-\frac{k_{1}-\sqrt{4g_2Dk_{2}/T+k_{1}^{2}}}{2 k_{2}}
\end{equation}
where $C_1$ is a constant determined by the initial condition.

\begin{equation}
    \label{RRA_C1}
    C_1 = \frac{k_{1} + \sqrt{4 g_2D k_{2}/T + k_{1}^{2}}}{k_{1} - \sqrt{4 g_2D k_{2}/T + k_{1}^{2}}}
\end{equation}

The second part is the solution of the equation which describes the reactions after irradiation ($t\geq T$). The solution is shown in \Eref{RRA_2_2}.
\begin{equation}
    \label{RRA_2_2}
    R(t)=\frac{k_1}{k_2}\frac{1}{C_2\exp(k_1t)-1}
\end{equation}
where $C_2$ is a constant determined by the concentration of peroxyl radical at the end of irradiation ($R(t)$), which can be calculated using \Eref{RRA_2_1}.
\begin{equation}
    \label{RRA_C2}
    \eqalign{C_2 =  e^{-Tk_1} 
    +\frac{ k_1^2Te^{- T k_1}}{2 g_2D k_2}\cr
    + \frac{(  e^{T \sqrt{4 g_2D k_2/T + k_1^2}} +  1) Tk_1e^{- T k_1} \sqrt{4 g_2D k_2/T + k_1^2}}{2 g_2 D k_2 (e^{T \sqrt{4 g_2D k_2/T + k_1^2}} - 1)} }
\end{equation}

Then, we calculate the integral of the solutions of these two parts based on \Eref{RA_AUC}. 
For the first part (integral of \Eref{RRA_2_1}), the integral interval is $[0, T)$, and the result is shown in \Eref{RRA_int1}.
\begin{equation}
    \label{RRA_int1}
    \eqalign{ 
    AUC[ROO\cdot]_{part1} =
    \frac{T (- k_{1} - \sqrt{4 g_2D k_{2}/T + k_{1}^{2}})}{2 k_{2}} - 
    \frac{\log{|-\frac{1}{C_1}+1|}}{k_2} \cr
    + \frac{\log{|-\frac{1}{C_1}+e^{T \sqrt{4 g_2D k_{2}/T + k_1^2}} |} }{k_{2}} 
    }
\end{equation}

For the second part (integral of \Eref{RRA_2_2}), the integral interval is $[T, +\infty)$. The result is shown in \Eref{RRA_int2}.
\begin{equation}
    \label{RRA_int2}
    AUC[ROO\cdot]_{part2} = \frac{k_{1} T}{k_{2}} - \frac{\log{|e^{k_{1} T} - \frac{1}{C_2} |}}{k_{2}}
\end{equation}

\begin{figure}
 \centering
    \includegraphics[scale=0.4]{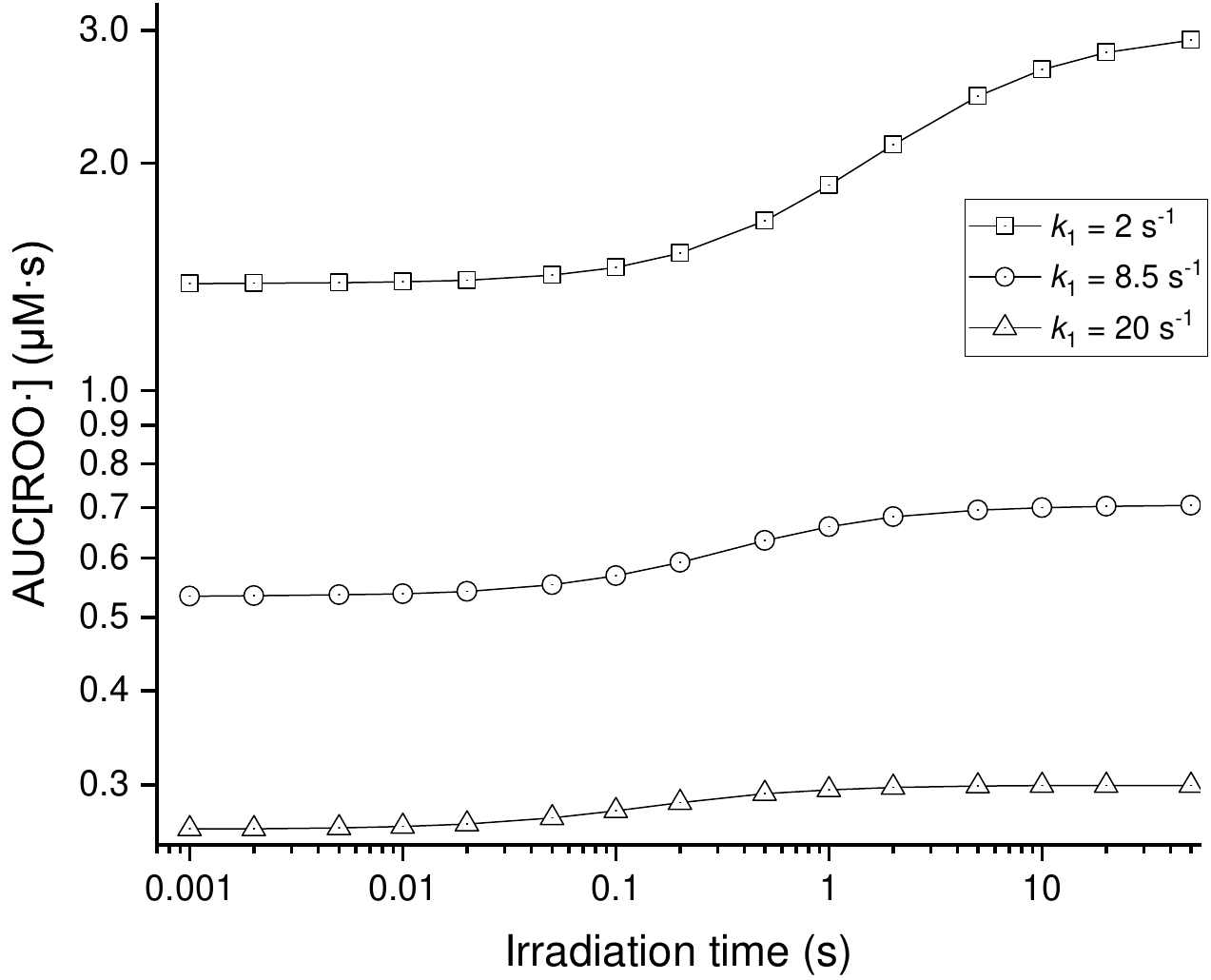}
    \caption{Concentration-time integrals of peroxyl radicals (AUC[ROO·]) for different irradiation times for tissues with different levels of antioxidants ($k_1$). The dose is 20 Gy. Each data point is an individual simulation.}
    \label{fig:RAT}
\end{figure}

The sum of these two parts (AUC[ROO·]) represents the damage related to peroxyl radicals. The magnitude of AUC[ROO·] against the total irradiation time, $T$, is shown in \fref{fig:RAT}. In this figure, each data point is an individual simulation, not the cumulative effect of results of different irradiation times.

The result indicates the time characteristic of the FLASH effect. The damage to normal tissue decreases with the decline of irradiation time, which forms an “S” shape curve. The steepest time interval of the curves is around hundreds of milliseconds corresponding to \citep{Montay-Gruel_2017}.
The “S” shape curve predicted by the hypothesis model indicates that the radical recombination-antioxidants hypothesis can explain the time characteristic of FLASH to a certain extent. 
The characteristic time is mainly determined by the reaction between peroxyl radicals and antioxidants, which is represented by the parameter $k_1$ in the model. As shown in the three curves with different $k_1$, the parameter can also greatly influence the difference in radiobiological effect between FLASH and CONV irradiation with the same dose delivery. 
The simulation results align with the radical recombination-antioxidants hypothesis, which deems that peroxyl radical-related reactions dominate the FLASH effect. It underscores that antioxidants compete to react with peroxyl radicals, which determines the time characteristic of the FLASH effect and the maximal difference between FLASH and CONV for a specific dose. This is due to the fact that the rates of these reactions are governed by the concentrations of the reacting species. The hypothesis infers that tissue with different antioxidant levels may exhibit different time characteristics of the FLASH effect \citep{hu_computational_2022}. The influence of $k_1$ on the time characteristic and difference between FLASH and CONV reflects this main point of the hypothesis.

\subsubsection{Radiobiological effects of FLASH and CONV irradiation}
\ 
\newline
\indent 
For a specific dose, D, the $DRF(T)$ (\Eref{DRF_RC}) can be derived by \Eref{RRA_int1}, \Eref{RRA_int2}, \Eref{RA_DAM} and \Eref{DMF_DEF} to estimate the biological effect brought by FLASH irradiation with different irradiation times. 

\begin{equation}
    \label{DRF_RC}
    DRF(T) 
    = \frac{f_1 (AUC[ROO\cdot]_{part1} + AUC[ROO\cdot]_{part2}) + f_2 D}
    {f_1 g_2 D /k_1 + f_2 D} 
\end{equation}

\begin{figure}
    \centering
    \subfigure[]{
    \includegraphics[scale=0.3]{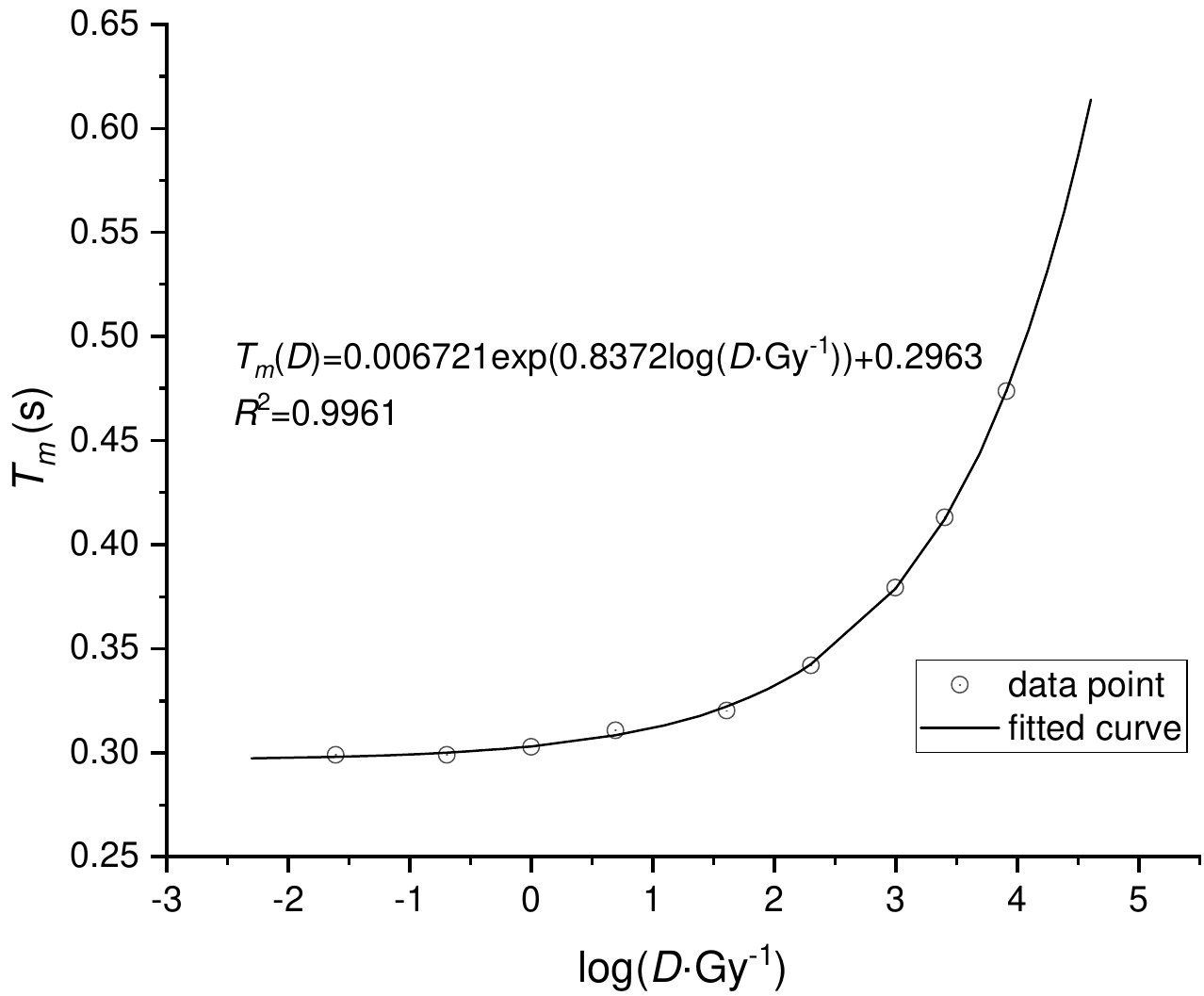}
    }
    \subfigure[]{
    \includegraphics[scale=0.3]{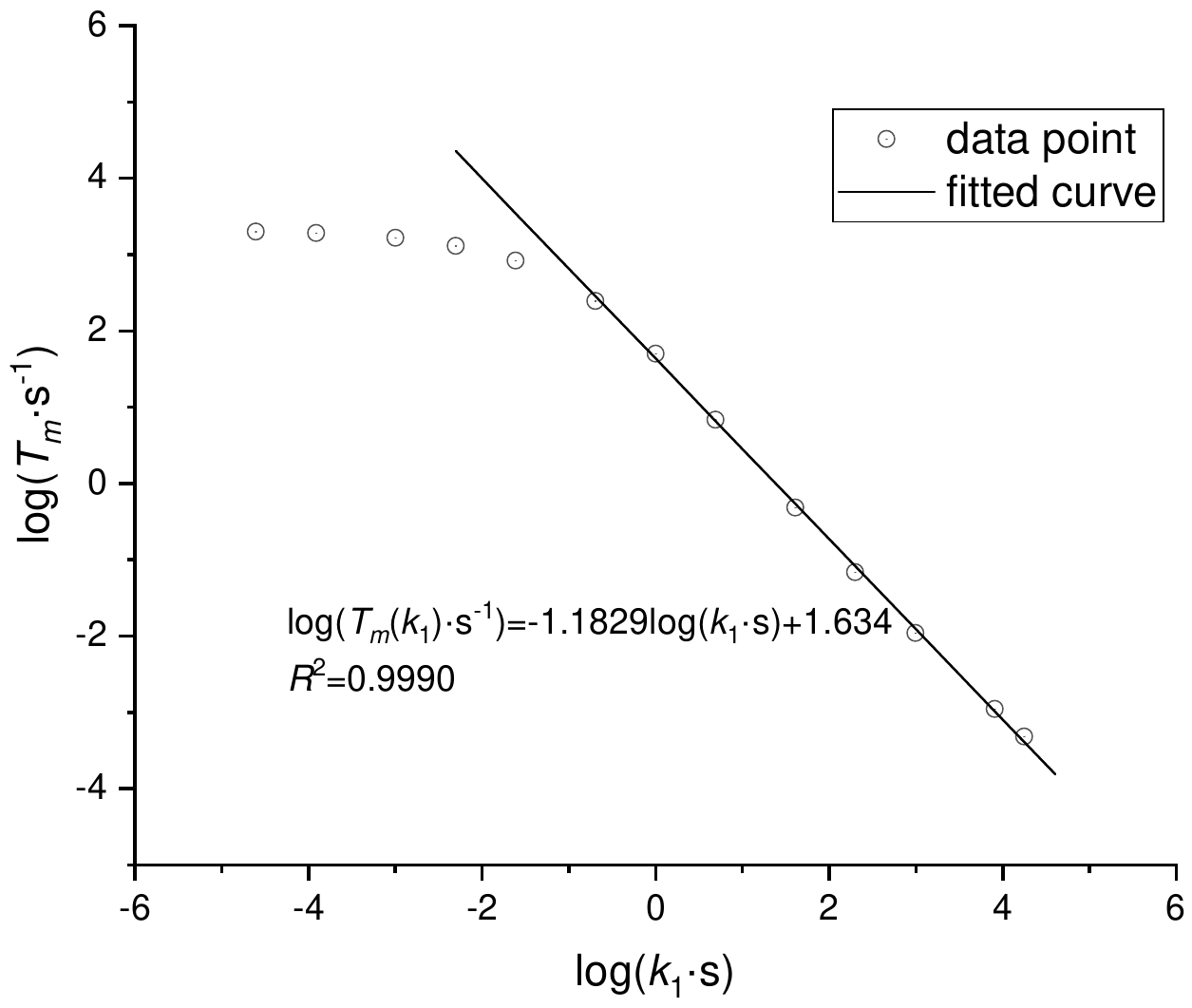}
    }
    \caption{Data points and fitted curves of the characteristic time, $T_m$, vs (a)Dose, $D$, and (b)antioxidant level, $k_1$}
    \label{fig:RRA_DMFT}
\end{figure}

However, the formula is too complex to explicitly reflect the relationship. 
We find a simple form in \Eref{RRA_DMFT} as an excellent approximation ($R^2>0.99$) of $DRF(T)$ in a wide range of parameters (we tested it by setting dose to 0.1-100 Gy and setting $k_1$ to 0.01-100 s$^{-1}$).
\begin{equation}
    \label{RRA_DMFT}
    DRF(T) = 1 - \frac{(1-DRF_{min})T_m}{T+T_m}
\end{equation}
where $DRF_{min}$ is the minimal value of $DRF$ for a given dose, which is equal to the value calculated by \Eref{RRA_DMF}; $T_m$ is a characteristic time, which is determined by $k_1$ and $D$. 
We also find that the formulas of $T_m(k_1)$ and $T_m(D)$ have concise approximate forms listed in \Eref{RRA_TM1} and \Eref{RRA_TM2}. The formula of $T_m(k_1)$ is obtained by excluding the data points where $k_1<1$, because the data points of $k_1>1$ can be approximated by a linear relationship on a log-log scale. If the radical recombination-antioxidants were the main mechanism of the FLASH effect, the value of $k_1$ should be higher than 1.0 s$^{-1}$ because it should explain the time characteristic of the FLASH effect observed by experiments \citep{Montay-Gruel_2017}. These formulas can be used as empirical formulas to predict the FLASH effect in practical application. 
\numparts
\begin{eqnarray}
\label{RRA_TM1}
T_m(D) = \varepsilon_1\exp{(\varepsilon_2\cdot \log{D} ) }+\varepsilon_3\\
T_m(k_1) = \vartheta_1\exp{(-\vartheta_2 \log{k_1})}, k_1\geq 1 \mathrm{s}^{-1}
\label{RRA_TM2}
\end{eqnarray}
\endnumparts
where $\varepsilon_1, \varepsilon_2, \varepsilon_3, \vartheta_1, \vartheta_2$ are parameters that can obtained by fitting. The data points used in fitting and fitted curves of $T_m(D)$ and $T_m(k_1)$ in \fref{fig:RRA_DMFT}. The results of curve fitting are listed in \tref{table:fit_RR}.

\begin{table}
\caption{Parameters and goodness of fitting curves of $T_m$ vs. $D$ and $k_1$}
\label{table:fit_RR}
\begin{tabular*}{\textwidth}{@{}l*{15}{@{\extracolsep{0pt plus12pt}}l}}
\br
Parameter & Value\\
\mr
$\varepsilon_1$ & 0.006721\\
$\varepsilon_2$ & 0.8372\\
$\varepsilon_3$ & 0.2963\\
$R^2$ for fitting $T_m(D)$ vs. $D$ & 0.9961\\
$\vartheta_1$ & 1.1829\\
$\vartheta_2$ & 1.634\\
$R^2$ for fitting $T_m(k_1)$ vs. $k_1$& 0.9990\\
\br
\end{tabular*}
\end{table}

For clinical application, the $T_m$ for a given dose can be obtained by fitting the data points of $(\log(T), DRF(D,T))$. The $T_m(D)$ can be obtained by fitting the data points from a series of doses.

Based on the methods list in section 2.2.3, we calculated the minimal damage induced by extremely short time ($T\to 0$) irradiation (\Eref{RRA_FLASH}) and CONV (\Eref{RRA_CONV}) irradiation to obtain the minimal value of $DRF$, i.e., $DRF_{min}$, to evaluate the maximal change of biological effect by a given dose of FLASH irradiation. 
\begin{equation}
    \label{RRA_FLASH}
    Damage_{min} = \frac{f_1\log (\frac{g_2k_2D}{k_1}+1)}{k_2} + f_2D
\end{equation}
\begin{equation}
    \label{RRA_CONV}
    Damage_{CONV} = \frac{f_1g_2D}{k_1} + f_2D
\end{equation}
Then the $DRF_{min}$ can be calculated and the result is shown in \Eref{RRA_DMF}.
\begin{equation}
    \label{RRA_DMF}
    DRF_{min}=\frac{Damage_{min}}{Damage_{CONV}} = \frac{[f_1 \log (\frac{g_2 k_2D}{k_1}+1)]/k_2+f_2D}{f_1g_2D/k_1+f_2D}
\end{equation}

\begin{figure}
    \centering
    \includegraphics[scale=0.4]{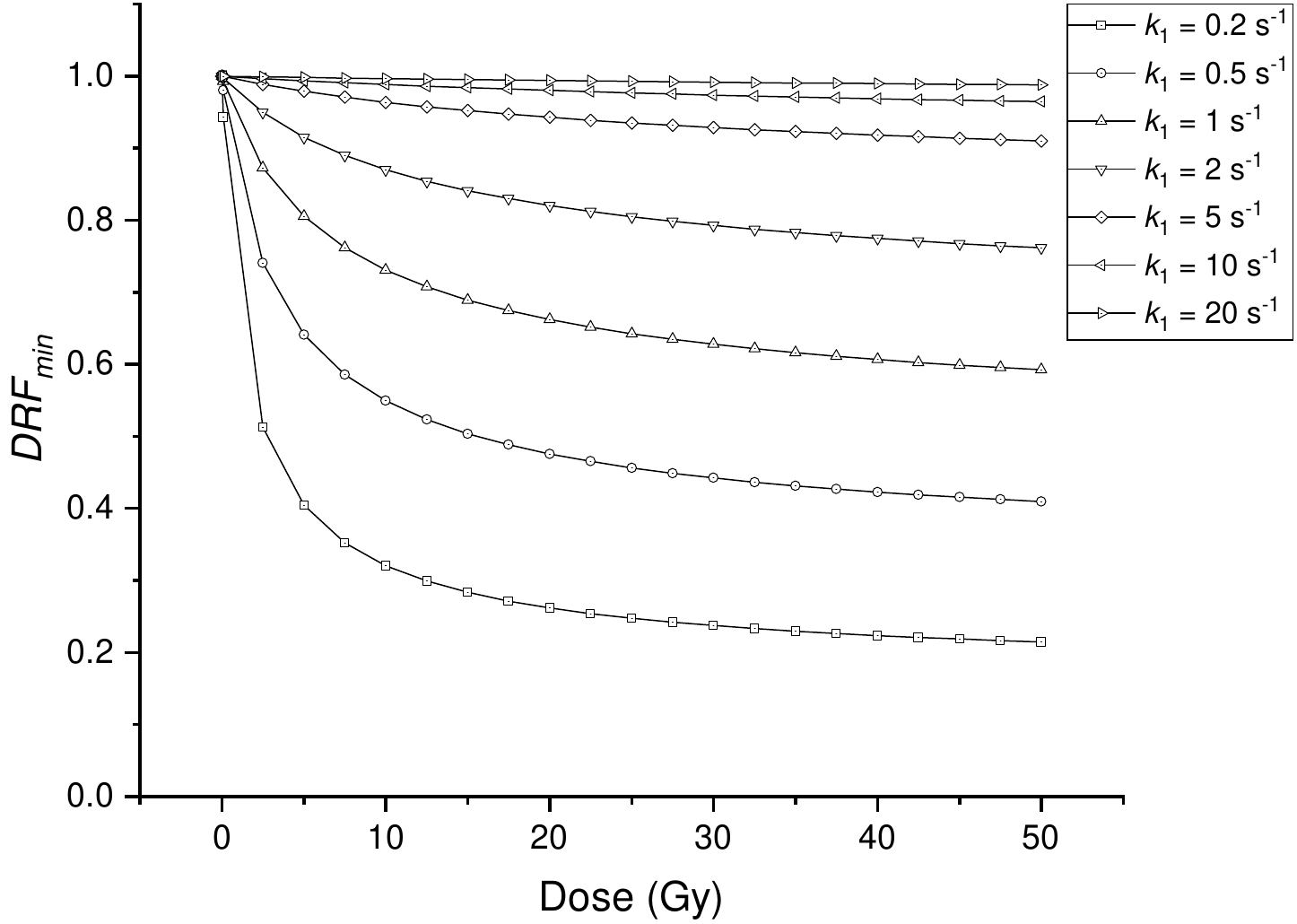}
    \caption{Minimal damage reduction ratio (irradiation time $T\to 0$), $DRF_{min}$, under different doses for cells with different levels of antioxidants, predicted by radical recombination-antioxidants hypothesis}
    \label{fig:DMFRA}
\end{figure}

We obtain a concise formula and it can be utilized to fit the experimental data conveniently. According to this formula, \Fref{fig:DMFRA} shows the $DRF_{min}$ irradiated by different doses under cells with different levels of antioxidants, $k_1$.

The curves indicate that the antioxidants in the cell greatly influence the change of biological effect attributed to the FLASH irradiation. 
This also corresponds to the hypothesis' explanation of the non-protective effect in tumors, which posits that the elevated antioxidant levels within tumors inhibit radical recombination, thereby narrowing the difference between the FLASH and CONV modalities. The value of $k_1$ for normal tissue is estimated to be 8.5 s$^{-1}$ \citep{hu_radical_2023} 
whereas in tumors, this value is several times higher.
Moreover, the curves show that the difference between FLASH and CONV irradiation increases with the increase of the total dose when the total dose is low. Then the $DRF_{min}$ changes relatively slightly with the increase of the total dose when the total dose is high enough. 

\subsection{Implementation of hypotheses for clinical applications}

Based on the mathematical modeling and analysis presented above, it can be observed that the FLASH effect is influenced by various key factors depending on the hypotheses. In this section, we elaborate on how to implement the hypothesis-based models for evaluating the efficacy of FLASH radiotherapy.

\subsubsection{Criteria for experimental data filtration}
\
\newline
\indent
The prediction of the FLASH effect based on the hypothesis-based model requires parameters from experiments. The control of variables is indeed essential in these experiments. However, it is hard to keep the consistency of all conditions because of the complexity of the biological system. Through the analysis of hypothesis-based models, it is feasible to ensure consistency in these key factors in the models within the same experimental group. Here, we will provide operational experimental conditions based on the key parameters in the two potential mechanism hypotheses, facilitating researchers to conduct systematic experiments and obtain reliable data for the assessment of FLASH effects.

For the oxygen depletion hypothesis, oxygen is the main point and the factors related to oxygen should remain the same. The density of capillaries and the types of cells in the tissue mainly determine the factors related to oxygen consumption and recovery, which are the main considerations in data filtration.
According to the oxygen depletion hypothesis, the criteria for the same group experiments are listed below.
\begin{enumerate}
    \item The tissues should have the same density of capillaries and the same tortuosity of cells for the same time characteristic of oxygen recovery, i.e., $\lambda$.
    \item The tissues should have the same metabolic oxygen consumption rate and the same blood supplement for the same initial oxygen concentration, i.e., $p_0$.
    \item The tissues should have the same level of other cellular compounds, i.e., $A$ in the model, competing to react with radiation-induced radicals alongside oxygen, among which cellular antioxidants are the most significant.
    \item The OER curves of the tissues should be similar, ensuring the same level of $m$ and $K$.
    \item The types of irradiation should be similar for the same yields of radicals, i.e., $g_1$.
\end{enumerate}

The radical recombination-antioxidants hypothesis mainly focuses on reactions of radicals and the level of antioxidants, not the oxygen in the tissue.
According to the radical recombination-antioxidants hypothesis, the criteria for the same group experiments are listed below.
\begin{enumerate}
    \item The tissues should have the same level of antioxidants, i.e., $k_1$.
    \item The responses of cells to irradiation should be similar for the same level of $f_1$ and $f_2$. 
    \item The types of irradiation should be similar for the same yields of radicals. It should be noted that the tissues should not be extremely hypoxic because low oxygen concentration could greatly compress the generation of peroxyl radicals, i.e., $g_2$.
\end{enumerate}

The experimental design or the retrospective study should consider the above criteria to avoid the influences on the conclusion brought by these factors.

\subsubsection{Suggestions for systematic experiments}
\ 
\newline
\indent
The protective effect of normal tissues is the most important biological advantage of FLASH effect, which can be quantified by the change of normal tissue complication probabilities (NTCP) induced by radiation at different dose rates. The NTCP models indicate that the NTCP curve often undergoes a substantial increase within a narrow dose interval \citep{palma_normal_2019}, beyond which it exhibits almost no variation. Notably, as an intrinsic characteristic of the NTCP, this narrow interval often shows at relatively high doses. Consequently, if an experiment aims to implement the change of NTCP as the endpoint of the FLASH effect, a single-dose irradiation cannot provide sufficient information for the effect at relatively low dose levels where the NTCP values still posit at low level for all dose rate setup and the differences cannot be observed. 
The existing data points on the FLASH effect primarily cluster at high doses \citep{bohlen_normal_2022,singers_sorensen_vivo_2022}, offering limited insights into the effect at relatively low dose levels, however, which are highly relevant for clinical radiotherapy and the verification of the hypotheses as well. 
To fill the gap, systematic experiments designed to explore the FLASH effect across a wide dose range are required. However, as for the low-dose level, it is challenging to obtain effective data points to represent the FLASH effect through simple experiments of FLASH irradiation in the one-fraction fashion, due to the intrinsic characteristic of the NTCP model.

To overcome the limitation, we propose a series of experiments using fractionated FLASH dose setups. In these experiments, the definition of fraction differs from the general definition in the field of radiotherapy. The duration between two consecutive fractions is several minutes rather than the common practice of one day. The total dose of FLASH irradiation is aimed to cover the high-dose interval where NTCP changes greatly, which is different from that of CONV irradiation because of the FLASH effect. However, for the limited granularity of fractionation dose, simple fractionated irradiation may not be able to achieve the total dose target. Thus, a hybrid irradiation strategy is proposed to supplement the limitation. The detailed description of the systematic experiments is listed below.
\begin{enumerate}
    \item \textbf{CONV group.} For CONV irradiation, the total dose is set to reach the interval where NTCP changes greatly. Then an endpoint is chosen to obtain the reference dose, $D_{CONV}$.
    \item \textbf{Fractionated FLASH dose setup.} For FLASH irradiation, set a series of   doses, $D_{FLASH,f}$, e.g., 1, 2, 5, 10,... Gy. For each setup of fractionated dose FLASH irradiation, we can define a virtual iso-effective dose, $D_{FLASH,ISO}(D_{FLASH,f})$, which induces the same radiobiological effect with the $D_{CONV}$ dose of CONV irradiation. The virtual iso-effective dose, $D_{FLASH,ISO}$, often cannot be obtained by fractionated FLASH irradiation solely because of the limited granularity of the fractionated dose.
    \item \textbf{Hybrid irradiation strategy.} For each dose setup of fractionated FLASH irradiation, to start with, $D_{FLASH,f}$ is delivered at FLASH dose rate for $n$ times to the experimental group with the duration of several minutes between two fractions, where $n$ is the maximal integer for $n\cdot D_{FLASH,f}\leq D_{FLASH,ISO}$. Then, the group is irradiated by a residual CONV dose, $D_{CONV,residual}$, to obtain the iso-effect dose for the fractionation dose setup, $D_{hybrid, ISO}(D_{FLASH,f})$, which induces the same radiobiological effect with $D_{CONV}$ CONV irradiation. The $D_{hybrid, ISO}(D_{FLASH,f})$ is the sum of $n\cdot D_{FLASH,f}$ and the residual CONV dose. 

With the systematic experiments, the DRF of the fractionation dose for FLASH effect can be calculated by \Eref{DMF_EX}.
\end{enumerate}
\begin{equation}
    \label{DMF_EX}
    \eqalign{
    &DRF(D_{FLASH,f}) \cr
    &= \frac{D_{CONV}-D_{CONV,residual}} {D_{hybrid,ISO}(D_{FLASH,f}) - D_{CONV,residual}} \cr
    &= \frac{D_{CONV}-\left[D_{hybrid,ISO}(D_{FLASH,f})-n\cdot D_{FLASH,f}\right]}{n\cdot D_{FLASH,f}}
    \cr
    &= 1 - \frac{D_{hybrid,ISO}(D_{FLASH,f})-D_{CONV}}{n \cdot D_{FLASH,f}}
    }
\end{equation}

These experiments enable the acquisition of $DRF$ versus dose and irradiation time curves across a wide range. The established dataset is expected to provide a fundamental base for advancing clinical FLASH radiotherapy. 

\subsubsection{Pipelines for obtaining parameters in models and predicting the FLASH effect}
\
\newline
\indent
Our work has established mathematical models based on the two potential hypotheses for the FLASH effect mechanism and analyzed the key factors respectively. Here, we give more concrete pipelines on how to implement the models into the clinical radiotherapy. Once any one of the hypotheses proves right, clinics can directly get the prediction of the FLASH effect for clinical cases according to the pipelines. 

It should be noted that the two mathematical models have different features, and the implementation pipelines are somehow different.

For the oxygen depletion hypothesis, the model parameters are generally accessible through measurements. They can be obtained by the steps listed below and then used to calculate the $DRF$ value by the analytical \Eref{DRF_ROD} for an exact dose,$D$, within the irradiation duration, $T$, and also the $DRF_min$ value by \Eref{RROD_DMF} for the extreme condition as $T\to 0$.

\begin{enumerate}
  \item Obtain the OER curve of the tissue to get the parameters, $m$ and $K$, within the curve.
  \item Measure the maximal radiolytic oxygen consumption irradiated by a series of given doses to obtain the yield of radiolytic oxygen consumption, $g_1$, and the equivalent concentration of cellular compounds, $A$, according to \Eref{ROD_dp}.
  \item Measure the oxygen recovery curve to obtain the feature constant, $\lambda$, or fit the data points with the same dose but a variety of irradiation times to obtain $\lambda$, fixing the parameters from the above two steps.
\end{enumerate}

With the parameters measured above, the FLASH effect can be predicted with the following steps in the clinical application.
\begin{enumerate}
    \item Calculate the $DRF_{min}$ vs. dose curve based on \Eref{RROD_DMF}. 
    \item Set a dose and calculate a series of $DRF$ for different irradiation durations according to \Eref{DRF_ROD}. 
    \item With the calculation results,the parameters, $\xi_1$ and $\xi_2$, in the approximate formula \Eref{RROD_FITBE} can be fitted. 
    It should be noted that these two parameters   are independent to doses and can be used for all doses because the variation of net oxygen concentration by the irradiation is independent to doses, as shown in \fref{fig:PTROD}.
    \item For a specific dose, $D$, one can use the approximate formula, \Eref{RROD_FITBE}, and the $DRF_{min}$  calculated from step 1 to calculate the $DRF$ vs. irradiation time curve and predict the FLASH effect for clinical use. 
   
\end{enumerate}

According to the radical recombination-antioxidants hypothesis, we can predict the FLASH effect solely based on the experimental data points of $DRF$ with different doses and irradiation times, other than the analytical calculations of $DRF$ and $DRF_{min}$ as above, because of the concise form of the approximate formula for $DRF$ (\Eref{RRA_DMFT}), 
and the inaccessibility of the model parameters, especially for the exact values of $f_1$ and $f_2$. 
The prediction steps are listed below.

\begin{enumerate}
\item Divide the experimental data points, $DRFs$, into several groups according to the dose.
\item For the data group with the same dose, fit the parameters $T_m$ and $DRF_{min}$ according to the form of \Eref{RRA_DMFT} using the experimental data points ($T$, $DRF$) classified in the first step.
\item Fit the curve of $DRF_{min}$ vs. dose as \Eref{RRA_DMF} using the data points ($D$, $DRF_{min}$) from the second step.
\item Fit the curve of $T_m$ vs. dose to obtain parameters in \Eref{RRA_TM1} using the data points ($D$, $T_m$) from the second step.
\item Predict the FLASH effect based on \Eref{RRA_DMF}, \Eref{RRA_TM1} and \Eref{RRA_DMFT} for a given dose and irradiation time.
\end{enumerate}

\subsection{Limitations}
\indent This study focuses on the quantitative analysis of the oxygen depletion hypothesis and the radical recombination-antioxidants hypothesis, which inevitably has some limitations.

Our work serves as a prospective mathematical modeling study to help quantitatively explore the FLASH mechanism. More importantly, it is an attempt to establish a potential FLASH effect prediction model for clinical application. 
Admittedly, the FLASH effect mechanism has not been unraveled currently and neither of the two hypotheses in this study has been confirmed reliable or widely accepted in the academic community. Both of them inevitably have some limitations and received some criticism \citep{Wardman_2022, hu_computational_2022}. 
In our study, we focus on the construction of the predictive model for the clinical application of the FLASH effect, and further quantitatively analyze how some key factors, e.g., initial oxygen, antioxidant level and total dose, determine the FLASH effect according to these two hypotheses. 
A series of experiments can be conducted with the results compared to our model predictions to validate the hypotheses. If one of the hypotheses model result is well in line with the experiments, our model can be used to quantitatively predict the FLASH effect for clinical application.

We use the amount of damage linear to the dose to evaluate the biological effect and calculate $DRF$. The definition of $DRF$ focuses on radiation-induced damage, which corresponds to the two hypotheses, rather than the cell's overall response to irradiation which is about to be depicted in the linear quadratic model. It should be noted that since the damage has not been clearly defined in the two hypotheses, the type of damage is not specified in this work.

It should be noted that real-world scenarios are highly simplified for modeling with some assumptions potentially deviating from actual conditions. The estimation of radiobiological effects relies on simplified models such as the oxygen enhancement ratio and the time integral of peroxyl radicals, which may not fully reflect the comprehensive impact. 

Other potential hypotheses of the FLASH mechanism are not included in this work, such as “protection of circulating immune cells”, “DNA integrity” and “mitochondrial damage response”, because these hypotheses lack quantitative descriptions. We cannot establish models to quantitatively link the key factors in these hypotheses to the FLASH effect. 

The existing FLASH experiments cannot provide sufficient data points to fit the parameters and achieve the prediction of the FLASH effect based on the models (seen in the section “Criteria for data filtration” and “Pipelines for obtaining parameters in models and predicting the FLASH effect”). Thus, we did not give examples of data fitting in the manuscript. When high-quality datasets are available in the future, the implementations of models can be achieved with the methods in our study.

The pipeline of obtaining parameters proposed in this study is a theoretical guideline. The practical methods to measure the required data (e.g. the measurement of OER curve for a tissue) are not included. With the development of FLASH radiotherapy and further study of the FLASH mechanism, efforts should be made to determine the key parameters in the prediction model of FLASH effect for specific patients.

\section{Conclusion}
\indent In this work, We formulated concise equations to abstract the oxygen depletion hypothesis and radical recombination-antioxidants into mathematical models. These equations were then solved to examine the influence of radiation features (total dose and irradiation time) and factors within the hypotheses (initial oxygen concentration and antioxidants) on the FLASH effect. Through the mathematical analysis, the formulas of $DRF$s are derived for clinical FLASH radiotherapy.

The mathematical analysis of these hypotheses highlights the key factors that determine the FLASH effect. In the case of the oxygen depletion hypothesis, the recovery of oxygen governs the timing feature of the effect, which is almost independent of doses. The competition between oxygen and other cellular compounds in reacting with radiation-induced radicals greatly impacts radiolytic oxygen consumption. The initial oxygen concentration plays a crucial role in the change of biological effects caused by FLASH irradiation. Notably, FLASH irradiation can greatly alter the biological effects of hypoxic and extremely hypoxic tissues.

In the case of the radical recombination-antioxidants hypothesis, the reaction between antioxidants and peroxyl radicals determines the timing characteristic of the FLASH effect. Antioxidants contribute to the differences in biological effects observed between FLASH and CONV irradiation. The $DRF$ exhibits a substantial increase with the total dose in the low-dose range, followed by a relatively slight change in the high-dose level.

The main implementation of the hypothesis in clinical radiotherapy is to predict the FLASH effect according to experimental data. Existing data is not sufficient to achieve this goal. We propose the criteria for data filtration according to the analysis of key factors influencing the FLASH effect. We provide suggestions for designing systematic experiments to obtain data in a wide dose range. These methods can provide datasets for subsequent data fitting and prediction. Then we describe the pipeline to fit parameters in hypothesis-derived models according to the solution to the equations. We show the methods to predict the FLASH effect in clinical applications based on the results of data fitting.

This study establishes a connection between the hypotheses of the FLASH mechanism and clinical radiotherapy by formulating predictive formulas for the FLASH effect utilizing parameters such as dose, irradiation time, and other biological factors. Furthermore, the mathematical analysis of key factors influencing the FLASH effect offers insights into the exploration of the FLASH mechanism.

\section*{Acknowledgments}
\indent This work was supported by the National Key Research and Development Program of China (Grant No. 2022YFC2402304 and 2021YFF0603600) and National Natural Science Foundation of China (Grant No. 12175114 and U2167209). The authors thank Kaiwen Li and Jianqiao Wang for their help in the preparation of the manuscript.

\bibliography{References}

\end{document}